\documentclass[12pt]{article}
\usepackage{epsfig}
\usepackage{graphicx,amsmath}
\def\Title#1{\begin{center} {\Large {\bf #1} } \end{center}}
\usepackage{fancyhdr}
\begin{document}
\topskip 2cm 

\Title{NLO automated tools for QCD and beyond}
\bigskip

\begin{raggedright}  

{\it \underline{Nikolas Kauer}  \index{}
\footnote{Presented at Linear Collider 2011: Understanding QCD at Linear Colliders  in searching for old and new physics, 12-16 September 2011, ECT*, Trento, Italy}\\
Department of Physics, Royal Holloway, University of London, Egham TW20 0EX, UK\\
{\rm n.kauer@rhul.ac.uk}
}

\bigskip\bigskip
\end{raggedright}
\vskip 0.5  cm
\begin{raggedright}
{\bf Abstract} 
Theoretical predictions for scattering processes with multi-particle final 
states at next-to-leading order (NLO) in perturbative QCD are 
essential to fully exploit the physics potential of present and future 
high-energy colliders.  The status of NLO QCD calculations and tools is 
reviewed.
\end{raggedright}


\section{Introduction}

The study of hard scattering processes at the Large Hadron Collider (LHC) 
\cite{Butterworth:2012fj} and a future TeV-scale linear collider is our primary 
means to probe and extend the Standard Model of particle physics.  It is 
driven by the comparison of experimental measurements with theoretical 
predictions, which depends on our ability to compute collider cross 
sections in perturbative QCD with adequate accuracy 
\cite{Salam:2011bj,Zanderighi:2012qr}.
This can only be achieved by going beyond leading order (LO) in QCD.
When using conventional measures, LO scale uncertainties are typically large 
compared to experimental uncertainties.  Moreover, for theoretical reasons 
a reliable estimation of the scale uncertainty is not feasible at LO.  
Consequently, an assessment of different scale choices, 
which is particularly important for many-particle/jet processes, 
is not possible.  Furthermore, the convergence of the perturbative series 
cannot be assessed at LO.  
When going beyond LO by including NLO corrections, the situation
improves significantly.\footnote{%
For processes with vastly differing scales, the resummation of
large logarithms of ratios of scales may also be necessary.}
At NLO, scale uncertainties can be assessed more reliably, and the 
residual uncertainties are often comparable to experimental uncertainties.\footnote{
Notable exceptions are the hadroproduction of Higgs and $Wb\bar{b}$ with $\sigma_\text{NLO}/\sigma_\text{LO}\approx 2$.}
NLO calculations thus deliver accurate predictions not only for the overall 
normalisation, but also for kinematic distributions including peripheral phase 
space regions.  This is in part due to the fact that new subprocesses often 
become active at NLO, which modify the normalisation and kinematic distributions.
Our ability to determine the uncertainty of parton distribution functions (PDF) and 
to model the structure of jets is also greatly enhanced at NLO.

In Section \ref{methods}, the state-of-the-art methods, implementations 
and tools for parton-level NLO calculations are briefly reviewed.  
In Section \ref{applications}, the status of collider physics applications 
is described.  The review ends with a summary.\footnote{%
The important topics of next-to-next-to-leading order (NNLO) calculations and 
combining parton-level fixed-order calculations and parton-shower event 
generators are beyond the scope of this review.}


\section{Methods, implementations and tools\label{methods}}

The structure and implied modularity of NLO calculations
is illustrated in Eqs.\ (1)--(3):
\begin{align}
\sigma_\text{NLO} =\ &\, \sigma_\text{Born} + \sigma_{corr} \\
\sigma_\text{Born} =\ & \int d\phi_n \;\frac{1}{2\hat{s}}\; 
\;\; |{\cal A}_{\text{LO}}|^2 \\
\sigma_{corr} =\ & \int d\phi_n \;\frac{\alpha_s}{2\hat{s}}\;
  \;\Bigl[ 
  \ \sum\limits_j\int d\phi_j  \mathcal{D}_j
+  {\cal A}_{\text{LO}}{\cal A}_{\text{NLO,V}}^* 
+ {\cal A}_{\text{LO}}^*{\cal A}_{\text{NLO,V}} \Bigr] \nonumber\\
& +\int d\phi_{n+1} \;\frac{\alpha_s}{2\hat{s}}\;
\;\,\Bigl[ |{\cal M}_{\text{NLO,R}}|^2 - \sum\limits_j \mathcal{D}_j \Bigr]
\end{align}
The new components of the NLO correction $\sigma_{corr}$ are:\footnote{%
The Born amplitude is assumed to be at tree level.}
the virtual corrections (involving one-loop amplitudes),
the real corrections (involving tree amplitudes)
and the infrared subtraction terms.\footnote{%
An alternative to the widely used subtraction formalism \cite{Ellis:1980wv} is the phase space slicing method \cite{nk_phase_space_slicing}.}
The resulting procedure for NLO calculations is
given in Table \ref{tab:NLOsteps}.
\begin{table}[htp]
\begin{center}
\fbox{
\begin{minipage}{0.85\textwidth}
1.\ Real correction: generate and evaluate $2\to n+1$ tree-level amplitudes\\
2.\ Subtract soft and collinear singularities due to single unresolved real radiation 
\phantom{2.\ }to obtain finite result\\
3.\ Integrate over $(n+1)$-particle phase space\\
4.\ Virtual correction: generate and evaluate UV-renormalised $2\to n$ one-loop 
\phantom{4.\ }amplitude after extraction of soft and collinear singularities to obtain finite 
\phantom{4.\ }result\\
5.\ Confirm cancellation of soft/collinear singularities (absorb initial state collinear 
\phantom{5.\ }singularities into PDF)\\
6.\ Integrate over $n$-particle phase space\\
7.\ Combine $2\to n+1$ and $2\to n$ contributions\\
8.\ Convolve with NLO PDF\\
9.\ Repeat for all contributing subprocesses
\end{minipage}
}
\end{center}
\caption{Steps to calculate the NLO QCD corrections for a $2\to n$ process. $n$ excludes electroweak decays.\label{tab:NLOsteps}}
\end{table}
The Binoth Les Houches Accord, a standard interface for combining the tree-level 
and loop-level contributions, has been defined in Ref.\ \cite{Binoth:2010xt} 
and is implemented in many automated tools (see below).

Until circa 2005, the limiting factor of NLO calculations 
was the computation of the virtual corrections, which
typically applied Passarino-Veltman (PV) \cite{nk_PV}
or PV-inspired \cite{nk_PV_inspired_reductions} tensor integral reduction
methods to evaluate the form factors of a Feynman-diagram-based amplitude
representation.  Several one-loop integral libraries are available as public codes:
LoopTools \cite{vanOldenborgh:1989wn,Hahn:1998yk},
QCDLoop \cite{Ellis:2007qk},
Golem95 \cite{nk_golem95},
OneLOop \cite{vanHameren:2010cp} and 
PJFry \cite{Fleischer:2010sq}.
The PV approach is general, but practical limitations arise due to
the factorial growth of the number of Feynman graphs with $N=n+2$,
the strong growth of the number of reduction terms with $N$ and due to
numerical instabilities for exceptional kinematic configurations,
which are caused by vanishing Gram determinants.
It has nevertheless been used successfully 
to create collections of NLO calculations 
based on analytic formulae and semi-automated methods, such as 
MCFM \cite{nk_MCFM,Campbell:2006xx}, 
MC@NLO \cite{nk_MCatNLO} and 
VBFNLO \cite{nk_vbfnlo,Hankele:2007sb,nk_vvjj_vbf,nk_waaj}.\footnote{%
The POWHEG BOX \cite{nk_powheg_method} library project \cite{nk_powheg_library,Jager:2011ms} 
was inspired by these collections.}
Since 2004, tremendous improvements have been achieved for the calculation of 
multi-leg one-loop amplitudes due to the exploitation of on-shell recursion 
relations and generalized-unitarity-cut constructibility 
as well as the possibility to even reconstruct the full rational terms 
\cite{nk_unitarity_methods,Berger:2008sj}.
On-shell reduction related tools are CutTools \cite{Ossola:2007ax}, Rocket \cite{nk_rocket} and Samurai \cite{Mastrolia:2010nb}.
Further innovative, complementary methods are also being developed 
\cite{nk_innovative}.
A comprehensive review of methods for multi-leg one-loop calculations can be 
found in Ref.\ \cite{Ellis:2011cr}.

Three widely-used algorithms for the generation of process-independent infrared subtraction terms are 
Catani-Seymour dipole subtraction \cite{nk_CS}, 
Frixione-Kunszt-Signer (FKS) subtraction \cite{nk_FKS} and 
antenna subtraction \cite{nk_antenna_subtraction}.\footnote{%
Research on alternative subtraction schemes is also being carried out \cite{Chung:2010fx}.}
Several implementations for these standard schemes are available:
Sherpa-Dipoles \cite{Gleisberg:2007md},
MadDipole \cite{nk_MadDipole},
HELAC-Dipoles \cite{Czakon:2009ss},
MadFKS \cite{Frederix:2009yq},
TeVJet \cite{Seymour:2008mu} and 
AutoDipole \cite{Hasegawa:2009tx}.

The following programs aim to provide
a comprehensive, automated solution for NLO 
calculations:
aMC@NLO \cite{Ossola:2007ax,Frederix:2009yq,nk_amcatnlo},
BlackHat/Sherpa \cite{Berger:2008sj,Gleisberg:2007md,Gleisberg:2008ta},
HELAC-NLO \cite{vanHameren:2010cp,Ossola:2007ax,Czakon:2009ss,nk_helac_nlo},
GoSam \cite{Cullen:2011ac},
FeynArts/FormCalc/LoopTools \cite{Hahn:1998yk,Hahn:2000kx} and 
MadGolem \cite{Binoth:2011xi}.


\section{Collider physics applications\label{applications}}

Discussions at the Les Houches 2005 Physics at 
TeV Colliders Workshop resulted in a list of processes for which the 
knowledge of NLO corrections was considered of particular importance
for the LHC physics programme \cite{Buttar:2006zd}.  This experimenter's NLO 
``wish list'' has guided theoretical efforts and was subsequently revised and updated 
in 2007\ \cite{Bern:2008ef} as well as 2009 \cite{Binoth:2010ra}.  The most recent
version is displayed in Table \ref{tab:wishlist}.
\begin{table}[htp]
  \begin{center}
{\footnotesize
     \begin{tabular}{|l|l|}
\hline \hline
Process ($V\in\{Z,W,\gamma\}$) & Comments\\
\hline
Calculations completed since Les Houches 2005&\\
\hline
&\\
1. $pp\to VV$+jet & $WW$+jet completed by
Dittmaier/Kallweit/Uwer~\cite{Dittmaier:2007th,Dittmaier:2009un};\\
 &
Campbell/Ellis/Zanderighi~\cite{Campbell:2007ev}.\\ 
 &
$ZZ$+jet completed by \\
&
Binoth/Gleisberg/Karg/Kauer/Sanguinetti~\cite{Binoth:2009wk}\\
2. $pp \to$ Higgs+2jets & NLO QCD to the $gg$ channel \\
& completed by Campbell/Ellis/Zanderighi~\cite{Campbell:2006xx};\\
& NLO QCD+EW to the VBF channel\\
& completed by Ciccolini/Denner/Dittmaier~\cite{Ciccolini:2007jr,Ciccolini:2007ec}\\
3. $pp\to V\,V\,V$ & $ZZZ$ completed 
by Lazopoulos/Melnikov/Petriello~\cite{Lazopoulos:2007ix}
 \\
 & and $WWZ$ by Hankele/Zeppenfeld~\cite{Hankele:2007sb}\\
 & (see also Binoth/Ossola/Papadopoulos/Pittau~\cite{Binoth:2008kt})  \\
4. $pp\to t\bar{t}\,b\bar{b}$ &  relevant for $t\bar{t}H$ computed by\\
 & Bredenstein/Denner/Dittmaier/Pozzorini~\cite{Bredenstein:2009aj,Bredenstein:2010rs} \\
 & and Bevilacqua/Czakon/Papadopoulos/Pittau/Worek~\cite{Bevilacqua:2009zn} \\
5. $pp \to V$+3jets & calculated by the Blackhat/Sherpa~\cite{Berger:2009ep} \\
 & and Rocket~\cite{Ellis:2009zw} collaborations\\
&\\
 \hline 
Calculations remaining from Les Houches 2005&\\
\hline
&\\
6. $pp\to t\bar{t}$+2jets & relevant for $t\bar{t}H$ computed by  \\
& Bevilacqua/Czakon/Papadopoulos/Worek
~\cite{Bevilacqua:2010ve}
\\ 
7. $pp\to VV\,b\bar{b}$,  & relevant for VBF $\rightarrow H\rightarrow VV$,~$t\bar{t}H$ \\
8. $pp\to VV$+2jets  & relevant for VBF $\rightarrow H\rightarrow VV$ \\
& VBF contributions calculated by \\
& (Bozzi/)J\"ager/Oleari/Zeppenfeld~\cite{nk_vvjj_vbf}
\\
\hline
NLO calculations added to list in 2007&\\
\hline
&\\
9. $pp\to b\bar{b}b\bar{b}$ & $q\bar{q}$  channel calculated by Golem collaboration~\cite{Binoth:2010pb} \\
&\\
\hline
NLO calculations added to list in 2009&\\
\hline
&\\
10. $pp \to V$+4jets & top pair production, various new physics signatures\\
11. $pp \to W b \bar{b}j$ & top, new physics signatures\\
12. $pp \to t\bar{t}t\bar{t}$ & various new physics signatures \\
&\\
\hline
Calculations beyond NLO added in 2007&\\
\hline
&\\
13. $gg\to W^*W^*$ ${\cal O}(\alpha^2\alpha_s^3)$& backgrounds to Higgs\\
14. NNLO $pp\to t\bar{t}$ & normalisation of a benchmark process\\
15. NNLO to VBF and $Z/\gamma$+jet  & Higgs couplings and SM benchmark\\
&\\
\hline 
Calculations including electroweak effects&\\
\hline
&\\
16. NNLO QCD+NLO EW for $W/Z$ & precision calculation of a SM benchmark\\
&\\
\hline
\hline
\end{tabular}
}
\end{center}
\vspace*{-0.4cm}
\caption{The experimenter's wish list for LHC processes in early 2010 (from \cite{Binoth:2010ra}).
\label{tab:wishlist}}
\end{table}

Due to the groundbreaking advances outlined in Section \ref{methods},
since 2009 the frontier for collider physics applications of NLO techniques
has also advanced considerably.
The following $2\to 4$ processes -- most are on the wish list -- have now
been calculated at NLO QCD:\footnote{%
$pp$ is given as initial state, but $p\bar{p}$ is also implied.}
$pp\to W\gamma\gamma+$jet \cite{nk_waaj},
$pp\to W+3$ jets \cite{Berger:2009ep,Ellis:2009zw,Berger:2009zg,KeithEllis:2009bu}, 
$pp\to Z,\gamma^\ast+3$ jets \cite{Berger:2010vm}, 
$pp\to t\bar{t}b\bar{b}$ \cite{Bredenstein:2009aj,Bredenstein:2010rs,Bevilacqua:2009zn,Bredenstein:2008zb},
$pp\to t\bar{t}jj$ \cite{Bevilacqua:2010ve,Bevilacqua:2011aa},
$pp\to b\bar{b}b\bar{b}$ \cite{nk_golem_bbbb},
$pp\to W^+W^-b\bar{b}$ \cite{Denner:2010jp},
$pp\to W^\pm W^\pm jj$ \cite{Jager:2011ms,Melia:2010bm},
$pp\to W^+W^-jj$ \cite{Melia:2011dw}
and most recently $pp\to$ 4 jets \cite{Bern:2011ep}.
Leptonic decays of weak bosons can be included trivially.
At the same level of complexity, complete off-shell effects for 
$pp\to t\bar{t}$ with dileptonic decay, 
i.e.\ $pp\to e^+\nu_eb\mu^-\bar{\nu}_\mu\bar{b}$, 
have been calculated at NLO QCD in Ref.\ \cite{Bevilacqua:2010qb},
which allowed to explicitly confirm the 
${\cal O}(\alpha_s\Gamma/M)$ effect predicted
by Ref.\ \cite{Fadin:1993dz}.
Advancing the frontier for linear collider physics, the process 
$e^+e^-\to 5$ jets has recently been calculated at NLO \cite{Frederix:2010ne},
which allowed to extract a competitive value of $\alpha_s(M_Z)$ from
5-jet LEP data.
Going beyond 4-particle final states in general requires the computation 
of 7-point one-loop amplitudes or higher.  This is the current complexity
frontier. At this level, NLO cross sections in leading-colour approximation
have been calculated for $V+4$ jets by the BlackHat/Sherpa collaboration 
($pp\to W+4$ jets \cite{Berger:2010zx} and $pp\to Z+4$ jets \cite{Ita:2011wn})
and for $e^+e^-\to n$ jets up to $n=7$ \cite{Becker:2011vg}.%
\footnote{%
Recently, the full-colour virtual contribution to $pp\to W+4$ jets has 
been calculated \cite{Ita:2011ar}.} 
The $n=7$ case required the computation of a one-loop 8-point function.


\section{Summary}

NLO QCD predictions for multi-particle processes are essential to fully exploit the 
physics potential of the LHC and a future linear collider.  In recent years, 
tremendous progress has been made in developing the calculational 
methods and tools that are required to compute NLO corrections for
hard scattering processes with 6, 7 or more external particles.  
At this level a (semi-)manual approach is no longer feasible, and the transition 
from collections of codes for specific processes to automated code 
generation for any process up to a maximum complexity has now been
achieved.  Several such automated tools are available or will become
public in the near future.  The modularity of NLO calculations allows 
to interface many tool components on the basis of the Binoth Les Houches Accord.


\section*{Acknowledgments}
I would like to thank the organisers for the invitation to speak
at Linear Collider 2011 and commend G.\ Pancheri and her team for 
hosting this well-organised and thoroughly enjoyable meeting.  
The hospitality of the European Centre for Theoretical Studies 
in Nuclear Physics and Related Areas (ECT*) as well as partial
support from ECT* and INFN are gratefully acknowledged.  
This work was carried out as part of the research programme of
the Royal Holloway and Sussex Particle Physics Theory Consortium
and the NExT Institute.
Financial support under the SEPnet Initiative from the Higher Education
Funding Council for England and the Science and Technology Facilities
Council (STFC) is gratefully acknowledged.  This work was supported by 
STFC grant ST/J000485/1.



\end{document}